\documentclass{article}
\usepackage{amssymb,graphicx,color}
\usepackage[table]{xcolor}

\def\Ai{{\rm Ai}}
\def\d{{\rm d}}
\def\mi{{\rm i}}
\def\P{{\rm P}}

\def\G{\Gamma}
\def\l{\lambda}
\def\z{\zeta}
\def\Re{\mathop{\rm Re\,}\nolimits}

\def\e{\mathop{\rm e}\nolimits}
\def\hf{{\textstyle{1 \over 2}}}

\def\qt{{\textstyle{1 \over 4}}}

\def\ni{\noindent}
\def\defi{\stackrel{\rm def}{=}}
\def\si{\!\!\! &}
\def\se{& \!\!\!}
\def\sm{\!\!\!}

\newcommand{\beq}{\begin{equation}}
\newcommand{\eeq}{\end{equation}}
\newcommand{\bea}{\begin{eqnarray}}
\newcommand{\eea}{\end{eqnarray}}

\definecolor{dgray}{gray}{.4}
\def\ccg{\cellcolor[gray]{0.85}}

\title{Exact sum rules for spectral zeta functions of\\
homogeneous 1D quantum oscillators, revisited}
\author{{\bf Andr\'e Voros}\\
Universit\'e Paris--Saclay, CNRS, CEA, Institut de Physique Th\'eorique\\
91191 Gif-sur-Yvette, France\\
E-mail: {\tt andre.voros@ipht.fr}\\[3pt]
\emph{dedicated to Professor Michael V. BERRY for his 80-th birthday}}

\begin{document}

\maketitle

\begin{abstract}

We survey sum rules for spectral zeta functions of homogeneous 1D 
Schr\"odinger operators, that mainly result from the exact WKB method.

\end{abstract}

It is a great honour and pleasure to write this note for Professor Michael Berry, 
to celebrate his major contributions to asymptotic analysis and wave mechanics,
particularly quantum and semiclassical theory, throughout his career.
I~have been very close to the semiclassical world since the age of 10 (Figure~1), 
and thus I have greatly benefited from his generous and inspiring guidance, 
warm encouragements and friendship in our numerous interactions. 
It all started with his major survey of semiclassical theory with K.~Mount, \cite{BM}
just when I began my PhD. 
Another highlight was when he offered me (with N.~Balazs and M.~Tabor)
to coauthor an article on quantum maps \cite{BBTV} which means very much to me;
and he inspired my activity throughout in exponential asymptotics
(with examples below) and in quantum chaology.
Indirectly, the group around him at Bristol and the whole British school of asymptotics 
have also proved very stimulating, and my gratitude goes to them as well.

\begin{figure}[h]
\center
\includegraphics[scale=.5]{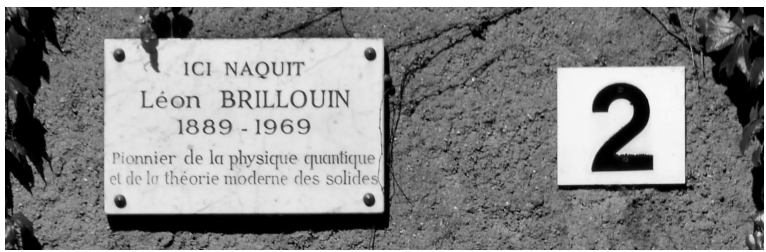}
\caption{\label{fg1} \small My walk to and from high school took me every day - unknowingly at the time -
alongside the house where L.~Brillouin was born, in the city of S\`evres, France.
}
\end{figure}

We present an overview of a specific early result in exact Wentzel--Kramers--Brillouin (WKB) analysis:
\emph{exact sum rules} for the \emph{spectral zeta functions} of homogeneous 1D 
Schr\"odinger operators, $\hat H_N = -\d^2/\d q^2 + |q|^N$ on $\mathbb R$
for ${N=1,2,\ldots}$
(Spectral zeta functions have been studied by Prof. Berry as well,
in actually harder 2D cases.~\cite{B})
Our sum rules generalize explicit formulae obeyed by the Riemann zeta function~$\z :$
\beq
\label{Z2M}
\z (2m) = \frac{(2\pi)^{2m} |B_{2m}|}{2 (2m)!}, \quad m=1,2,\ldots \qquad
(B_{2m}: \mbox{ Bernoulli numbers}). 
\eeq
Indeed, as we will recall in \S~2.1, (\ref{Z2M}) is a case of spectral sum rules for one such operator:
%The Riemann $\z $ function is indeed (up to an elementary factor) a spectral zeta function: that of
the harmonic oscillator $\hat H_2$, for which the spectrum is also exactly solved 
by the basic semiclassical WKB method. 
This is no coincidence: the extension of (\ref{Z2M}) to other homogeneous cases is likewise a by-product
of exact WKB handling generalized to anharmonic potentials. \cite{V0}\cite{VA}
We have much detailed the case of the quartic potential $q^4$ before \cite[\S~10]{V1};
here we sketch a more global picture disclosing some original exact identities: 
(\ref{ZN2})--(\ref{Z35}), among which (\ref{Z6P2}) and (\ref{Z3+2}) have \emph{fully closed forms 
quite similar to} (\ref{Z2M}). We will skip all but minimal background and details, pointing to references instead.

\section{The potentials $|q|^N$ and exact WKB analysis}

\subsection{Spectral functions, and notations}

The (parity-symmetric) Schr\"odinger operator $\hat H_N$ is known
to have a purely discrete positive energy spectrum $\{ E_k \}_{k=0,1,\ldots}$ ($N$-dependence is implied),
with $E_k \uparrow + \infty$, and $E_k \propto k^{2N/(N+2)}$ asymptotically, for $k \to +\infty$.
It then admits full, resp. parity-twisted, \emph{spectral generalized-zeta functions} \cite[\S~1]{VA}
\beq
Z_N (s,\l ) \defi \sum_k (E_k +\l )^{-s}, \quad \mbox{resp. } Z_N^\P (s,\l ) \defi \sum_k (-1)^k(E_k +\l )^{-s} 
\eeq
(the sums converge for $\Re s > \frac{N+2}{2N}$); and definite-parity variants as well,
\beq
Z_N ^\pm (s,\l ) \defi \sum_{k\ {\scriptstyle{\rm even}\atop \scriptstyle{\rm odd}}}  (E_k +\l )^{-s} 
= \hf [Z_N \pm Z_N^\P] (s,\l ) 
\eeq
($+/-$ amounts to imposing a Neumann/Dirichlet condition at the center ${q=0}$).

We write $Z_N^\#$ for any of $(Z_N,Z_N^\P,Z_N^+,Z_N^-)$ (and likewise for other kinds of spectral functions).
Either pair $(Z_N,Z_N^\P)$ or $(Z_N^+,Z_N^-)$ suffices to generate the quartet, 
but formulae may favour any of the four types, so we keep them all.

The functions $Z_N^\# (s,\l )$ are \emph{meromorphic in the whole $s$-plane and regular at} $s=0$.
For us they only serve as a gateway to two other (sets of) spectral functions in one variable each,
\bea
D_N^\# (\l ) \si\defi\se \exp [-\partial_s{Z_N^\#}(s,\l ) |_{s=0} ]
\qquad \qquad \ \mbox{(spectral determinants)} , \\
Z_N^\# (s) \si\defi\se Z_N^\# (s,\l =0) \qquad \qquad \qquad \qquad \mbox{(spectral zeta functions)} .
\eea
The latter zeta functions go back to \cite[\S~7]{MP} in general, and to \cite{P} for our $|q|^N$ case.
They enter these fundamental expansion formulae for the determinants:
\beq
\label{DTZ}
\log D_N^\# (\l ) = - {Z_N^\#}'(0) - \sum_{n=1}^\infty \frac{{Z_N^\#}(n)}{n} (-\l )^n .
\eeq
The radius of convergence here is the lowest relevant eigenvalue, always $>0$.
And the values $Z_N(1), \ Z_N^\pm (1)$ for $N=1,\ 2$ (for which the defining series diverge) 
are meant as regularized values or finite parts.

Being homogeneous, $\hat H_N$ is invariant modulo dilations under complex rotations
$q \mapsto \e^{\mi \varphi _N} q$ with $\varphi_N \defi 4\pi /(N+2)$;
the order $L_N$ of that symmetry (the lowest integer $L>0$ making 
$L \varphi _N \equiv 0 \bmod 2\pi $) is
\beq
\label{LN}
L_N= \hf N+1 \ \mbox{ for $N$ even}, \qquad N+2 \ \mbox{ for $N$ odd}.
\eeq
The degree $N$ will most often occur through the combination
\beq
\nu _N \defi 1/(N+2) = \varphi _N/(4\pi ) \qquad 
\mbox{(to be denoted $\nu $ for maximum brevity)}.
\eeq

\subsection{Exact WKB approaches}

Exact WKB analysis uses the idea that for 1D potentials analytic in the complex $q$-plane,
the ordinary complex-WKB equations for the Schr\"odinger solution can be reinterpreted 
to describe its analytical continuation in $q$, an operation that gives \emph{fully exact} results.
The topic much grew around 1980, but quite a few earlier works we listed in \cite[\S~1.2]{Ve} were crucial to it; 
we owe Prof. Berry our awareness of an essential building block, Dingle's treatment of asymptotics. \cite{Di} 
It is much beyond the scope of this note (limited to a special early application)
to go into the overall subject - which now has several variants, and we know of no truly global reference either.

Our own use of an exact WKB scheme was in part triggered by the Berry--Mount survey quoted above,
as it very clearly exposed the complex WKB method and highlighted its limitations: 
thus, for the simplest (i.e., parabolic) quantum 1D barrier problem,
even that most efficient semiclassical method left the \emph{phase} of the transmission coefficient 
\emph{undetermined}. \cite[\S~3.3]{BM}
We were at the time aiming that complex WKB method at the quartic oscillator $\hat H_4$, \cite{BPV} 
but the above issue was definitely blocking us and had to be cleared first.
The upshot was that upon using \emph{complex} $\hbar$, and borrowing from Balian--Bloch, \cite{BB} Sato--Kawai--Kashiwara, \cite{SKK} and Dingle, \cite{Di}
the complex WKB method does describe the full transmission coefficient (or Jost function) \emph{exactly}
for a quadratic barrier \cite[\S~7]{V1} - an approach which then at once extended 
to a quartic potential and beyond. 

The exact-WKB \emph{method} we used went by fully describing ramified analytic Borel transforms of spectral determinants.
But exact-WKB \emph{results} were at once twofold, connected by Borel--Laplace \emph{equivalence}:
in the Borel plane, we had exact but quite abstract \emph{discontinuity formulae} \cite{V1} 
(soon recognized as instances of ``bridge equations" in \'Ecalle's \emph{resurgence} framework); \cite{E} 
and on the spectral determinants, upon Borel resummation, those gave exact \emph{functional relations}, \cite{V0} 
akin to those of Sibuya for Stokes multipliers, \cite[chap.~5]{S} 
and later reduced to one equivalent bilinear equation: \cite[eq.~(5.32)]{Vs}\cite[\S~IV.3]{VM}
\bea
\label{WE}
\e^{\mi \nu \pi } D_N^+ (\lambda) D_N^-(\e^{4\mi \nu \pi } \lambda)
\!-\! \e^{-\mi \nu \pi } D_N^+ (\e^{4\mi \nu \pi } \lambda) D_N^- (\lambda) 
\si=\se 2 \mi \qquad \qquad (N \ne 2) \\
\label{WE2}
\si=\se 2 \mi \e ^{-\mi\pi\l /4} \ \ (N = 2). \qquad
\eea
Enforcing the compatibility of such functional relations with the expansions (\ref{DTZ}) readily generated
concrete \emph{exact polynomial identities} for the zeta-values $Z_N'(0),Z_N^\#(n)$ $(n=1,2,\ldots )$: \cite{V0}
those are the early \emph{spectral sum rules} to be surveyed here.

Later, the form (\ref{WE}) further led to \emph{exact quantization conditions} \cite{VM}\cite{VA} 
for which we see \emph{no Borel-plane equivalent} at all (see \S~3.3);
then, (\ref{WE}) was also linked to the Bethe Ansatz: the ``ODE/IM correspondence" \cite{DT}\cite{BLZ}.
Thus, \emph{direct} exact-WKB analysis (upon the spectral functions themselves) 
began to look more far-reaching than \emph{resurgent} exact-WKB analysis (still needed to attain (\ref{WE}), however).
And finally, a wholly direct derivation of (\ref{WE}) arose (reviewed in \cite{Ve}\cite{VD}) 
that made the lengthy detour through the Borel plane redundant. 
So for our spectral sum rules, resurgent or direct exact-WKB analysis can do;
overall however, a direct approach currently seems more complete. 
(But we don't imply the resurgent framework to be obsolete, especially for nonlinear problems!)

As other exact-WKB schemes, we may think of Silverstone's work with uniform WKB methods, \cite{Su} 
of hyperasymptotic analysis, \cite{BH} and of nonlinear extensions for, e.g., Painlev\'e functions. \cite{KT}

%In the end, the primary concrete output of an exact WKB analysis of the operators $\hat H_N$ 
%is an exact bilinear functional equation in the style of Sibuya \cite[chap.~5]{S} 
%constraining the spectral determinants $D_N^\pm$: \cite{V0}\cite{VM}\cite{VA}
\section{Special known cases}

\subsection{The harmonic oscillator ($N=2$)}

$\hat H_2$ has the explicit spectrum $\{E_k = 2k+1 \}_{k=0,1,\ldots}$
giving the spectral determinants
\beq
\label{Gam}
\textstyle D_2(\l ) = 2^{-\l /2} \sqrt{2\pi}/\G \bigl( \frac{1+\l }{2} \bigr) ,
\qquad D_2^\P (\l ) = 2 \, \G (\frac{3+\l }{4})/ \G (\frac{1+\l }{4}) ,
\eeq
and spectral zeta functions which are the Dirichlet lambda and beta functions respectively: \cite[eqs.~(23.2.20--21)]{AS}
\beq
\label{ZZP}
Z_2 (s) = (1-2^{-s}) \z (s), \qquad Z_2^\P (s) = \beta (s) .
\eeq

The functional equation (\ref{WE2}) obeyed here by the determinants amounts to 
the \emph{reflection formula for the Gamma function} and can be split as \cite[\S~4.1]{Ve}
\beq
D_2 (\l ) D_2 (-\l ) = 2 \cos \hf \pi \l , \qquad
D_2^\P (\l ) / D_2^\P (-\l ) = \cot \qt \pi (1 - \l ).
\eeq
In order to make use of (\ref{DTZ}), we take the logarithms and expand 
around $\l =0$, then identify the resulting sides order by order.

With $D_2$ we thus get
\beq
2 \biggl[ -{Z_2}'(0) - \sum_{m=1}^\infty \frac{Z_2 (2m)}{2m} \l ^{2m} \biggr]\equiv
\log 2 - \sum_{m=1}^\infty \frac{|G_{2m}|}{4m (2m)!} (\pi \l )^{2m} ,
\eeq
with the right-hand side using the suitable primitive of a classic Taylor formula,
$\tan \hf \pi \l = \sum\limits_{m=1}^\infty \frac{\textstyle |G_{2m}|}{\textstyle (2m)!} (\pi \l )^{2m-1}$ 
($G_{2m}$: Genocchi numbers). \cite{O}

With $D_2^\P$ we likewise get
\beq
2 \sum_{m=0}^\infty \frac{Z_2^\P (2m+1)}{2m+1} \l ^{2m+1} \equiv
\sum_{m=0}^\infty \frac{|E_{2m}|}{(2m+1)!} (\hf \pi \l )^{2m+1} :
\eeq
the right-hand side now came by integration from
\ $\sec \hf \pi \l = \sum\limits_{m=0}^\infty \frac{\textstyle |E_{2m}|}{\textstyle (2m)!} (\hf \pi \l )^{2m}$ ($E_{2m}$: Euler numbers). 
Overall, we have thus reached: ${{Z_2}'(0) = -\hf \log 2}$, and \cite[chap.~23]{AS}
\beq
\label{ZP2}
Z_2 (2m) = \frac{\pi^{2m}}{4(2m)!} |G_{2m}| \quad (m>0), \qquad
Z_2^\P (2m+1) = \frac{(\hf \pi)^{2m+1}}{2(2m)!} |E_{2m}| .
\eeq
Now the identities: $G_{2m} \equiv 2(1-2^{2m})B_{2m}$, \cite{O} and (\ref{ZZP}) for $Z_2$, together make
the formula (\ref{ZP2}) for $Z_2 (2m)$ equivalent to (\ref{Z2M}) for $\z (2m)$ with Bernoulli numbers.
But (\ref{ZP2}) with Genocchi numbers is more \emph{symmetric}: 
the $G_{2m}$ are \emph{integers} $(-1,1,-3,17,\ldots)$, just like the $E_{2m}$.
Moreover, we deliberately drew (\ref{ZP2}) (or (\ref{Z2M})) straight from the Gamma-function reflection formula: 
not only this approach is most elementary, but it specifically extends to anharmonic cases.

\subsection{The case $N=1$}

The spectral determinants for the potential $|q|$ use the Airy function, \cite[\S~3]{VA}
\beq
D_1^+(\l )=-2\sqrt \pi \Ai '(\l ), \qquad D_1^-(\l )=2\sqrt \pi \Ai (\l ).
\eeq
A special property here, not seen for spectral determinants $D_N^\#$ with $N>1$,
is that $\Ai (\l )$ obeys a \emph{linear differential equation} (Airy's equation), 
making all its Taylor coefficients at 0 explicit: 
${\rm Ai}^{(n)}(0) = 3^{(n-2)/3} \pi ^{-1} \sin \frac{2}{3}(n+1)\pi \, \G (\frac{n+1}{3})$.
\cite[chap.~10.4]{AS}
Then, $\log D_1^\pm$ can in turn be expanded to any order in closed form, giving
\bea
\label{ZAM}
{Z_1^+}'(0) \si=\se %\textstyle \log \frac{3^{1/3} \G(1/3)}{2\sqrt \pi}, \quad
\hf \log [\sqrt 3 / (2 \rho )] , \quad 
{Z_1^-}'(0)= %\log \frac{3^{2/3} \G(2/3)}{2\sqrt \pi}
\hf \log [\sqrt 3 \, \rho /2]
\quad \mbox{ (regularized sums)} , \qquad \\ 
Z_1^+(1) \si=\se 0, \qquad \qquad \qquad \quad Z_1^-(1) = -\rho 
\qquad \qquad \quad \ \ (\qquad " \qquad \qquad " \ ), \\
Z_1^+(2) \si=\se 1/\rho, \qquad \qquad \quad \quad Z_1^-(2)=\rho ^2, \\ 
\label{ZAP}
Z_1^+(3) \si=\se 1, \qquad \qquad \qquad \quad Z_1^-(3)= \hf -\rho ^3, \\[-4pt]
\si\vdots\se \qquad \qquad \qquad \qquad \qquad \quad \, \vdots \nonumber
\eea
\beq
\label{RO}
\mbox{with } \rho \defi D_1^\P (0) = -\Ai'(0)/\Ai(0) 
= 3^{5/6}(2\pi)^{-1} \G(2/3)^2 \ (\approx 0.729011133).
\eeq

Here the main exact-WKB functional equation (\ref{WE}) reduces to 
a classic Wronskian relation \cite[eq.~(10.4.12)]{AS}. 

The spectral zeta function $Z_1^-$ (aka Airy zeta function) is covered in earlier literature,
\cite[\S~4]{Cr}\cite[\S~1.11 ex.~48, 50]{BBG}
but not its partner $Z_1^+$ until \cite{VA}, to our knowledge.
In particular we find no previous mention of the exceptional rational value
${Z_1^+(3)=1}$ (the inverse cubes of the unsigned zeros of $\Ai'$ sum up to unity!).

\section{The general 1D homogeneous case}

\subsection{General results (for integer degrees $N>0$)}

For general potentials $|q|^N$, the extension of the Gamma-function reflection formula
from the case $N=2$ is thus the functional relation (\ref{WE})--(\ref{WE2}).
If we now expand it in powers $\l ^n$ using (\ref{DTZ}) then identify sides order by order, 
we attain a countable sequence of \emph{exact identities} or \emph{sum rules}:
\bea
\label{ZD0}
(n=0:) \qquad \qquad \qquad \qquad \qquad \quad {Z'_N}(0) \si=\se \textstyle \log \sin \nu \pi \\
\label{ZZ1}
- \cot \nu \pi \sin 2 \nu \pi \, Z_N^\P (1) + \cos 2 \nu \pi \, Z_N(1) 
\si=\se 0 \quad [\mbox{indeterminacy for } N=2] \qquad \ \\
\label{ZZ2}
- \cot \nu \pi \sin 4 \nu \pi \, Z_N^\P (2) + \cos 4 \nu \pi \, Z_N(2) 
\si=\se - 4 \cos^2 \nu \pi \, Z_N^\P (1)^2 \\
\label{ZZ3}
- \cot \nu \pi \sin 6 \nu \pi \, Z_N^\P (3) + \cos 6 \nu \pi \, Z_N(3) 
\si=\se 4 \cos^2 \nu \pi \bigl[ 2 \cos^2 \nu \pi \, Z_N^\P (1)^3 \nonumber\\
&& \ {} - 3 \cos 2 \nu \pi \, Z_N^\P (1) Z_N^\P (2) \bigr] \\[-8pt]
\si\vdots\se \nonumber\\[-2pt]
\label{ZZn}
- \cot \nu \pi \sin 2n \nu \pi \, Z_N^\P (n) + \cos 2n \nu \pi \, Z_N(n) 
\si=\se {\mathcal P}_{N,n} \{Z_N^\P(m)\}_{1 \le m<n}\\[-4pt]
\si\vdots\se \nonumber
\eea
with ${\mathcal P}_{N,n}$ being homogeneous polynomials of degree $n$ if $Z_N^\P(m)$ are assigned degree $m$.
(A path from (\ref{WE}) to (\ref{ZD0})--(\ref{ZZn}) is best detailed in \cite[\S~IV.2--3]{VM}.)

For $N=2$, (\ref{ZZn}) expresses $Z_2(n)$ for $n>1$ even and $Z_2^\P(n)$ for $n$ odd, like (\ref{ZP2}).
So, (\ref{ZZn}) does generalize the harmonic-case result (\ref{ZP2}) - but with two shortcomings:

a) the left-hand side of (\ref{ZZn}) is not (scaling to) a basic zeta-value 
$Z_N^\#$ (i.e., $Z_N$ or $Z_N^\P$ or $Z_N^\pm$) in general, 
but is rather one weird linear combination thereof - in which the weights recur with $n$-period $L_N$ given by (\ref{LN});

b) the right-hand side in (\ref{ZZn}) is not explicit in general, because hardly any values $\{ Z_N^\P(m) \}_{m<n}$
get determined through (\ref{ZZn}) recursively: only one linear combination of $\{Z_N^\#(n)\}$
gains an expression at each $n$, when two independent ones would be needed. 

To mitigate a), the left-hand side of (\ref{ZZn}) \emph{sometimes} scales to a basic $Z_N^\#(n)$:

- $Z_N(n)$ if $n=\ell L_N$ for $\ell$ integer (except: $Z'_N(0)$ for $n=0$);

- $Z_N^\P(n)$ if $N \equiv 2 \bmod{4}$ and $n= \hf \ell L_N$ for $\ell$ odd;

- $Z_N^\pm(n)$ if $N$ is odd and $n=(\ell +\hf ) L_N \mp \hf $ for $\ell$ integer.

To mitigate b), \emph{all values for $n=0$ and $1$} are actually known \emph{in closed form}, 
by earlier, not exact-WKB, arguments: \cite{Vz} 
\beq
\label{Z0}
n=0 : \quad (Z_N^\P)'(0)= \log \bigl[ \nu^{N\nu } \G (\nu )/\G(1-\nu ) \bigr] , \quad 
Z'_N (0) = \log \sin \nu \pi \qquad \quad
\eeq
(curiously, the exact-WKB method recovers $Z'_N(0)$ as (\ref{ZD0}), but not at all ${Z_N^\P}'(0)$);
\beq
\label{Z1}
n=1 : \quad Z_N^\P (1) = 
%\frac{\sin \nu \pi}{2 \sqrt\pi} (2\nu)^{2N \nu} \frac{\G(\nu) \G(2\nu) \G(3\nu)}{\G(2\nu+\hf)} ,
\frac{\sqrt\pi}{2} (2\nu)^{2N \nu} \frac{\G(2\nu) \G(3\nu)}{\G (1 -\nu) \G(2\nu +\hf)} ,
\quad Z_N(1) = \frac{\tan 2\nu\pi}{\tan \nu\pi} \, Z_N^{\rm P}(1) 
\eeq
(by Weber--Schafheitlin integral formulae; the exact-WKB method still misses $Z_N^\P$ at $n=1$, 
but it does catch the \emph{ratio} $Z_N (1)/Z_N^\P (1)$ this time, through (\ref{ZZ1})).
Remark: (\ref{Z0})--(\ref{Z1}), stated in \cite{Vz} with $N$ even, hold for $N$ odd as well. \cite[\S~3]{Cr}

(The results (\ref{ZD0})--(\ref{Z1}) do not fall into the same class as earlier (also exact) $Z$-value formulae
but at the \emph{nonpositive} integers $n=0,-1, \ldots$: \emph{trace identities}. \cite{P}
The latter arise through expanding the partition function $\sum_k \e^{-tE_k}$ for $t \to 0^+$, 
an operation wholly belonging to conventional (not exact) asymptotic analysis. 
$Z$-values at \emph{positive} integers have an unrelated, more transcendental, nature.
For the Riemann $\z $ function, $\z (-n)$ and $\z (1+n)$ are related by the Riemann Functional Equation, 
but \emph{this} is a special arithmetic situation.)

\subsection{Examples}

For convenience, Tables~\ref{T1}--\ref{T2} list the closed-form expressions (\ref{Z0})--(\ref{Z1}) 
and the exact (possibly rescaled) identities (\ref{ZD0})--(\ref{ZZ3}) for low values of $N$
- splitting even vs odd $N$ because the parameter $L_N$ given by (\ref{LN}) will reenter.

Tables~\ref{T3}--\ref{T4} then present a synoptic view of the exact spectral identities 
(always up to rescalings) according to their left-hand sides which they evaluate.

Both Tables \ref{T3}--\ref{T4} highlight as \emph{unshaded} the cells with values 
we know in \emph{fully explicit} closed form. That includes:
the leftmost column in each Table, from (\ref{ZP2}) and (\ref{ZAM})--(\ref{ZAP}) respectively
(here, exact-WKB sum rules alone determine only half the data for $N=2$, and $Z_1'(0)$ for $N=1$);
then the top 2 rows ($n=0,1$) from (\ref{Z0})--(\ref{Z1}), determining complete data ($Z_N$ and $Z_N^\P$). 
Further closed forms fill the third rows ($n=2$): these are now specifically \emph{exact-WKB results} 
from the sum rule (\ref{ZZ2}) helped by (\ref{Z1}), but that only covers one specific linear combination 
of $Z_N(2)$ and $Z_N^\P(2)$ at each~$N$,
\beq
\label{ZN2}
\cot \nu \pi \sin 4 \nu \pi \, Z_N^\P (2) - \cos 4 \nu \pi \, Z_N(2) =
\frac{\pi (2\nu )^{4N\nu }}{4} \Biggl[ \frac{\G(\nu ) \G (3\nu )}{\G(1 \!-\! 2\nu ) \G(2\nu \!+\! \hf )} \Biggr]^2 .
\eeq
($Z_N(n)$, $Z_N^\P(n)$ values taken separately are known only 
in terms of generalized hypergeometric $\vphantom{F}_4 F_3$ infinite series for $n=2$,
and of $n$-uple definite integrals for general~$n$: \cite[App.~C]{V1} 
neither of those do we call fully closed forms.)

Two, and only two, exceptional basic spectral-zeta values $Z_N^\# (2)$
then lie among those new ($N>2$) closed forms (\ref{ZN2}), 
thus creating the closest anharmonic analogs of (\ref{ZP2}), namely (\ref{Z6P2}) and (\ref{Z3+2}) below:

- with $N$ even (Table \ref{T3}), not for the quartic, but for the \emph{sextic} oscillator: 
(\ref{ZZ2}) and (\ref{ZN2}) reduce to
\beq
\label{Z6P2}
Z_6^\P (2) = \sqrt 2 \, Z_6^\P (1)^2 = \textstyle 
\frac{1}{8} \, [\pi \, \G (\frac{5}{4})]^2/\G (\frac{7}{8})^4 \ (\approx 0.71895230),
\eeq
which allows (\ref{ZZ3}) to add one exceptional closed-form result for $n=3$,
\bea
(1+\sqrt 2) \, Z_6^\P (3) + Z_6(3) \si=\se %(3 \sqrt 2 +2) \, Z_6^\P (1) ^3 \nonumber\\
-(3\sqrt 2 + 4) \, Z_6^\P(1)^3 + 3(2 + \sqrt 2) \, Z_6^\P(1)Z_6^\P(2) \nonumber\\
\label{Z4E}
\si=\se \textstyle (3 +\sqrt 2) \, 2^{-19/4} [\pi \, \G (\frac{5}{4})]^3 /\G (\frac{7}{8})^6 
\ (\approx 2.26279887). \qquad
\eea
The next identity, for $n=4$, also original, can be cast (see Note after (\ref{Z35})) as
\beq
\label{Z64}
Z_6(4) = \textstyle \frac{1}{3}(248-175 \sqrt 2) \, Z_6(1)^4 - \frac{4}{3}(2-\sqrt 2) \, Z_6(1)Z_6(3) 
\eeq
(but no longer in fully closed form, because we do not know $Z_6(3)$ analytically).
For completeness, we also specify the last marked ($N=6$)-cell in Table \ref{T3}:
\bea
\label{Z6P6}
Z_6^\P (6) \si=\se \textstyle -\frac{1}{30} (210+151 \sqrt 2) \, Z_6^\P (1)^6 
+\hf (34+23 \sqrt 2) \, Z_6^\P (1)^4 Z_6^\P (2) \nonumber\\
\si\se \textstyle {}-\hf (18+15 \sqrt 2) \, Z_6^\P (1)^2 Z_6^\P (2)^2  
+\hf (2+ \sqrt 2) \, Z_6^\P (2)^3 \nonumber\\
\si\se \textstyle {}-\frac{2}{3} (6+5 \sqrt 2) \, Z_6^\P (1)^3 Z_6^\P (3) 
+2(2+ \sqrt 2) \, Z_6^\P (1) Z_6^\P (2) Z_6^\P (3) \nonumber\\
\si\se \textstyle {}-\frac{1}{3} \sqrt 2 \, Z_6^\P (3)^2 
+\frac{6}{5} \sqrt 2 \, Z_6^\P (1) Z_6^\P (5)
\eea
Numerical confirmations for (\ref{Z6P2})--(\ref{Z6P6}) follow from \cite[Table]{V0}
(or \cite[\S~10 and Table~5]{V1}, but together with the \emph{errata} on the quoted Web page).

- with $N$ odd (Table \ref{T4}), for the \emph{cubic} oscillator: \cite[chap.~5]{S}\cite[\S~3]{VA} 
using the golden ratio $\phi = \hf (1 + \sqrt 5)$, (\ref{ZZ2}) and (\ref{ZN2}) reduce to
\beq
\label{Z3+2}
Z_3^+ (2) = \phi \, Z_3^\P (1)^2 = \textstyle 
(\frac{2}{5})^{2/5} \phi ^{-1} \pi \, [\G (\frac{6}{5})/\G (\frac{9}{10})]^2
\ (\approx 0.993522181) ,
\eeq
but nothing for $Z_3^- (2),\ Z_3^\P (2)$; hence
we have no closed form for the $(n=3)$ cell in Table \ref{T2} (i.e., for $Z_3^-(3)$),
to match (\ref{Z4E}) from the sextic case.
%as we have no sum rule to determine $Z_3^\P (2)$.
The next identity for a basic ($N=3$) zeta-value (Table \ref{T4}), also new, can be written as
\bea
\label{Z35}
Z_3(5) \si=\se \textstyle \frac{1}{48} (369163-165095 \sqrt 5)\, Z_3(1)^5
+ \frac{5}{24} (2503-1119 \sqrt 5) \, Z_3(1)^3 Z_3(2) \nonumber\\
&& \textstyle {} + \frac{5}{16} (23-11\sqrt 5) \, Z_3(1)Z_3(2)^2
+ \frac{5}{6} (-31+14 \sqrt 5) \, Z_3(1)^2 Z_3(3) \nonumber \\
&& \textstyle -\frac{5}{6} \, Z_3(2) Z_3(3) + \frac{5}{8} (-7+3\sqrt 5) \, Z_3(1) Z_3(4)
\quad (\approx 0.8949120) .
\eea
(For numerical checks in the cubic case: see Appendix.)

Note: (\ref{Z64}) and (\ref{Z35}) exemplify the general fact that when the left-hand side of (\ref{ZZn})
is $\propto Z_N(n)$ ($\Leftrightarrow n$ is a multiple of $L_N$), 
then the right-hand side too is expressible in $Z_N$-values alone;
that is because the full determinant $D_N(\l )$ inherits from (\ref{WE}) 
an \emph{autonomous} functional equation with a symmetry of order~$L_N$. \cite{VA} 
(For $N=1$ and 4, both having $L_N=3$, $Z_1(3)$ is recalled that way in Table \ref{T2}, 
$Z_4(3)$ in Table \ref{T1}, and $Z_4(6)$ is given by \cite[eq.~(6)]{V0}.)

Suggestion (for experts in PT-symmetry):  could their homogeneous non-selfadjoint potentials,
such as $\mi q^3$, yield similar spectral sum rules as well?

All examples above very concretely testify to the capacity of the WKB framework 
to yield \emph{analytically exact results} for anharmonic oscillators.

\subsection{Concluding remarks}

As further results, spectral sum rules have been found for 
more general potentials \cite{St}\cite{Cr}, 2D billiards \cite{IML}\cite{B} 
and compact hyperbolic surfaces (Selberg-zeta zeros), \cite[App.~B]{VB}
plus analogous sum rules for the Riemann zeros as well. 
\cite[Tables]{VB}

Now, the early exact-WKB results we have reviewed don't close the story;
fortunately so, because they appear heavily underdetermined
(one functional equation (\ref{WE}) for two unknown functions $D_N^\pm$, 
one sum rule (\ref{ZZn}) for two independent $Z_N$-values at each $n$).
But as detailed in \cite{VM}\cite{VA}, obvious a priori constraints on $D_N^\pm$ 
(that they are entire functions with all their zeros \emph{real negative} and \emph{asymptotically known})
suffice to obtain (from (\ref{WE}) alone) separate \emph{fixed-point equations} for $D_N^+$ and $D_N^-$;  
those equations have moreover been \emph{proved} to be \emph{contracting} in this homogeneous-potential case: \cite{A}
thereby, both $D_N^\pm$ end up \emph{fully determined},
which makes \emph{this} direct exact-WKB approach achieve an exact and effective quantization of the spectrum. 
Now:

\ni - all that extra information beyond (\ref{WE}) seems of a more global nature,
not reducible to another set of finite spectral sum rules that would complete (\ref{ZZn})
(save for $N=1$, as (\ref{ZAM})--(\ref{ZAP}) may suggest);

\ni - and we currently have no idea whether and how those crucial extra properties and results
could emerge in a purely Borel-plane resurgent approach.

Finally, exact WKB treatments also work to solve general (inhomogeneous) polynomial potentials
(as reviewed in \cite{Ve}, and more lately \cite{VD}).
\smallskip

{\bf Acknowledgment}: we used Mathematica \cite{W} to attain 
(\ref{Z64}), (\ref{Z6P6}),~(\ref{Z35}).

\section*{Appendix: numerical data for the cubic case}

Computer evaluations of the functions $Z_3^\#$ support our new spectral sum rules for the potential $|q|^3$.

First, (\ref{Z3+2}) can be checked to arbitrary precision with, say, Mathematica, \cite{W} using (\ref{Z1}), 
\beq
\label{Z31}
Z_3^\P (1) \approx 0.7836009674833 , \qquad Z_3(1) \approx 3.319386965494 ,
\eeq
followed by \cite[eqs.~(C.27--28) and (C.32--33)]{V1} which respectively reduce to
\bea
Z_3(2) \si=\se Z_3^\P (1)^2 + \biggl ( \frac{2}{5} \Bigr )^{2/5} \biggl [ \frac{5-\sqrt{5}}{8 \pi} \biggr ] ^{1/2}
\frac{ \G (\frac{3}{5}) \G (\frac{4}{5}) }{ \G (\frac{13}{10})} \,
{}_4F_3 \biggl( {\frac{4}{10} \atop}\, {\frac{5}{10} \atop \frac{12}{10}}\, {\frac{6}{10} \atop \frac{13}{10}}
\, {1 \atop \frac{14}{10}} ; 1 \biggr) \nonumber \\
\label{Z32}
\si\approx\se 1.098003371 \\
\label{Z3-2}
\mbox{and } Z_3^-(2) \si=\se \Bigl ( \frac{2}{5} \Bigr )^{7/5} 
\frac{ \G (\frac{7}{10}) \G (\frac{4}{5}) }{3 \sqrt \pi \, \G (\frac{7}{5})} \,
{}_4F_3 \biggl( {\frac{6}{10} \atop}\, {\frac{7}{10} \atop \frac{14}{10}}\, {\frac{8}{10} \atop \frac{15}{10}}
\, {1 \atop \frac{16}{10}} ; 1 \biggr) \approx 0.104481190 ,
\eea
upon which $Z_3^+(2) = Z_3(2) - Z_3^-(2)$ leads to full numerical agreement with (\ref{Z3+2}).

Then, the displayed identities involving $n>2$: that for $Z_3^-(3)$ in Table \ref{T2}, and (\ref{Z35}) for $Z_3(5)$, 
can be tested using an Euler--Maclaurin numerical approximation for the sum defining $Z_N(s)$
(and a similar one for the alternating sum $Z_N^\P(s)$),~\cite{Vz}
\beq
\label{EM}
Z_N(s) \sim \sum_{k<K} E_k^{\, -s} + \hf E_K^{\, -s} + \frac{\mu b_0}{2\pi} \frac{E_K^{\, -s+\mu }}{s-\mu }
+ \Bigl [ \frac{-\mu b_1}{2\pi (s \!+\! \mu )} + \frac{B_2}{2} \frac{2\pi }{\mu b_0} s \Bigr ] E_K^{\, -s-\mu }
\eeq
($\!\! \bmod {{\rm O}(E_K^{\, -s-3\mu }})$), in terms of $\mu \defi (N+2)/(2N)$ and of 
the first Bohr--Sommerfeld expansion coefficients $b_j$, 
which for $N=3$ read as $b_0=\frac{2^{2/3}}{5} \sqrt 3 \, \G(\frac{1}{3})^3 /\pi $, 
$b_1=-\frac{2^{4/3}}{9} \pi ^2 / \G(\frac{1}{3})^3$. \cite[eq.~(7.18)]{V1}
Using the eigenvalues for $k \le K=9$ supplied in \cite[eq.~(22)]{VA}, 
(\ref{EM}) yields an \emph{estimated} accuracy $\gtrsim 6$ digits for $n \ge 3$:
\beq
Z_3^-(3) \approx 0.025878, \quad Z_3(3) \approx 0.9646441, \quad Z_3(4) \approx 0.9210896 ;
%\quad Z_3(5) \approx 0.8949120
\eeq
for $n \le 2$ that accuracy drops below 5 digits, but we may then switch back to (\ref{Z1}), (\ref{Z32}), (\ref{Z3-2}) 
which allowed us arbitrary accuracy.

\begin{table}[h]
\center
\vskip -5mm

\caption{\label{T1} \small For $N$ even, top 4 rows ($n=0,\ 1$): closed-form values from (\ref{Z0})--(\ref{Z1});
bottom 4~rows ($n=0$ to 3): exact-WKB sum rules, from (\ref{ZD0})--(\ref{ZZ3}); 
the two sets share the middle 2 rows.}
\vskip -5pt

\begin{tabular} {cccc}
\hline\\[-10pt]
\hfill $N : \sm $ & 2 & 4 & 6 \\
\hline\\[-8pt]
$\sm {Z_N^\P}'(0) \sm$ & 
$\!\! \log [2^{-3/2} \G (\qt )^2 /\pi ]$ 
& $\sm\sm \log [2^{-7/3} 3^{1/3} \G (\frac{1}{3})^4 /\pi ^2] $ 
& $\log [2^{-9/4} \G (\frac{1}{8}) / \G (\frac{7}{8})] $ \\[4pt]
$Z_N^\P (1)$ & $\pi /4$ 
& $2^{-11/3} 3^{-1/3} \G (\frac{1}{3})^5 /\pi ^2$ & 
$2^{-7/4} \, \pi \, \G (\frac{5}{4}) /\G(\frac{7}{8})^2$ \\[4pt]
\hline\\[-8pt]
$Z_N'(0)$ & $-\hf \log 2$ & $-\log 2$ 
& $\hf \log [(2 \!-\! \sqrt 2)/4]$ \\[3pt]
$Z_N(1)$ & $\infty $ & $3 \, Z_4^\P (1)$ 
& $(1 \!+\! \sqrt 2 ) \, Z_6^\P (1)$ \\[3pt]
\hline\\[-8pt]
$n=2$ & $\sm Z_2(2)=2Z_2^\P(1)^2 \sm $ 
& $\sm 3Z_4^\P(2) \!+\! Z_4(2)=6Z_4^\P(1)^2 \sm $
& $Z_6^\P(2)=\sqrt 2 \, Z_6^\P(1)^2$ \\[4pt]
$n=3 $ & $Z_2^\P(3)=$ \hfill & $Z_4(3)= \qquad \qquad \qquad \quad \ $ \hfill 
& $\sm (\sqrt 2 \!+\! 1) Z_6^\P(3) \!+\! Z_6(3)= $ \hfill \\[2pt]
& \hfill $2Z_2^\P(1)^3$ & $\sm \frac{9}{2}[-Z_4^\P(1)^3 \!+\! Z_4^\P(1)Z_4^\P(2)] \sm $
& \hfill $-(3\sqrt 2 \!+\! 4)Z_6^\P(1)^3 + $ \\[2pt]
& & \hfill $=\frac{1}{6} Z_4(1)^3 - \hf Z_4(1)Z_4(2)$ & \hfill $\sm 3(2 \!+\! \sqrt 2)Z_6^\P(1)Z_6^\P(2)$ \\[4pt]
\hline
\end{tabular}
\bigskip

\caption{\label{T2} \small As Table \ref{T1} but for $N$ odd. In the $N=1$ column: $\rho $ as in (\ref{RO}); 
in the $N=3$ column: $\phi =\hf (\sqrt 5+1)$, the golden ratio.}
\vskip 4pt

\begin{tabular} {ccc}
\hline\\[-10pt]
\hfill $N : $ & 1 & 3 \\
\hline\\[-8pt]
${Z_N^\P}'(0)$ & $-\log \rho$ 
& $\log [5^{-3/5} \G (\frac{1}{5}) /\G (\frac{4}{5})]$ \\[4pt]
$Z_N^\P (1)$ & $\rho $ 
& $(\frac{2}{5})^{1/5} \phi ^{-1} \sqrt \pi \, \G (\frac{6}{5}) / \G(\frac{9}{10})$
\\[4pt]
\hline\\[-8pt]
$Z_N'(0)$ & $\log [\sqrt 3 /2] $ & $\hf \log [(5 \!-\! \sqrt 5)/8] $ \\[3pt]
$Z_N(1)$ & $-Z_1^\P (1) $ & $(2 \!+\! \sqrt 5) \, Z_3^\P (1)$ \\[3pt]
\hline\\[-8pt]
$n=2$ & $Z_1^-(2)=Z_1^\P(1)^2$ & $Z_3^+(2) =\phi \, Z_3^\P(1)^2$ \\[4pt]
$n=3$ & $Z_1(3)= \hf Z_1^\P(1)^3 + \frac{3}{2} Z_1^\P(1)Z_1^\P(2)$ 
& $Z_3^-(3)= \qquad \qquad \qquad \qquad \qquad $ \hfill \\[2pt]
& $\qquad \, = \frac{5}{2} Z_1(1)^3 - \frac{3}{2} Z_1(1)Z_1(2)$
& $-(\phi \!+\! \hf) Z_3^\P(1)^3 + \frac{3}{2} Z_3^\P(1)Z_3^\P(2)$ \\[4pt]
\hline
\end{tabular}
\end{table}

\begin{table}[h]
\center
\caption{\label{T3} \small In black: the left-hand sides in the exact-WKB sum rules from (\ref{ZD0})--(\ref{ZZn}) 
for $N$ even; $\ast$ stands for generic forms.
(Save for the $'$ signs in the top row, each $N$-column has a pattern of $n$-period
$L_N=\hf N+1$ as given by (\ref{LN})).
In grey: additional exact data, from (\ref{Z0})--(\ref{Z1}). 
In shaded cells: data we \emph{cannot} cast in fully closed form.}
\vskip 4pt

\begin{tabular} {lllllll}
\hline \\[-10pt]
$n \!\bigm \backslash\! N \!\!\!$ & \ \ 2 & \ 4  & \ 6  & \ 8  & \ 10  & $\ \cdots$ \\
\hline \\[-8pt]
$0$ & $Z'_2 \color{dgray} {,Z_2^\#}'$ & $\!\! Z'_4 \color{dgray} {,Z_4^\#}'$ 
& $\!\! Z'_6 \color{dgray} {,Z_6^\#}'$ & $\!\! Z'_8 \color{dgray} {,Z_8^\#}'$ 
& $\!\!\!\! Z'_{10} \color{dgray} {,Z_{10}^\#}'$ & $\ \cdots$ \\[8pt]
$1$ & $Z_2^\P$ & \multicolumn{4}{c}{$\longleftarrow $
\qquad \quad $\ast {\color{dgray} ,Z_N^\#} \qquad \quad \longrightarrow$} & $\ \cdots$ \\[11pt]
$2$ & $Z_2$  &  $\ \ast$  & $Z_6^\P$ & $\ \ast$  &  $\ \ast$  & $\ \, \ast$ \\[12pt]
$3$ & $Z_2^\P$ & \ccg $Z_4$ &  $\ \ast$ & \ccg $\ \ast$  & \ccg $\!\! Z_{10}^\P$ &\ccg $\ \, \ast$ \\
$4$ & $Z_2$  &  \ccg  $\ \ast$  & \ccg $Z_6$  & \ccg $\ \ast$   & \ccg  $\ \ast$  
& \ccg \raisebox{6pt}{\rotatebox[origin=c]{20}{$\ddots$}} \\[12pt]
$5$ & $Z_2^\P$ & \ccg  $\ \ast$  &  \ccg  $\ \ast$   & \ccg $Z_8$ & \ccg  $\ \ast$  & \ccg \\[12pt]
$6$ & $Z_2$  & \ccg $Z_4$ & \ccg $Z_6^\P$ & \ccg  $\ \ast$ & \ccg $\!\! Z_{10}$  & \ccg \\[10pt]
$\,\vdots$ & $\ \vdots$ &\ccg $\ \vdots$ \quad \rotatebox[origin=c]{-15}{$\! \ddots$} 
&\ccg $\ \vdots$ \quad \rotatebox[origin=c]{-15}{$\! \ddots$} &\ccg $\ \vdots$ &\ccg $\ \vdots$ & \ccg $\ddots$ \\[1pt]
\hline
\end{tabular}
%\end{table}
\smallskip

%\begin{table}[h]
%\center
\caption{\label{T4} \small As with Table 1, but for $N$ odd hence now $L_N=N+2$ by (\ref{LN}), 
and adding exact data from (\ref{ZAM})--(\ref{ZAP}) in the $N=1$ column (also in grey).}
\vskip 4pt

\begin{tabular} {lllllll}
\hline \\[-10pt]
$n \!\bigm \backslash\! N \!\!\!$ & \hfill 1 \hfill & \ 3  & \ 5  & \ 7  & \ 9  & $\!\! \cdots$ \\
\hline \\[-8pt]
$0$ & $Z'_1 \color{dgray} {,Z_1^\#}'$ & $\!\! Z'_3 \color{dgray} {,Z_3^\#}'$ 
& $\!\! Z'_5 \color{dgray} {,Z_5^\#}'$ & $\!\! Z'_7 \color{dgray} {,Z_7^\#}'$ 
& $\!\! Z'_9 \color{dgray} {,Z_9^\#}'$ & $\!\! \cdots$ \\[8pt]
$1$ & $Z_1^+\color{dgray} ,Z_1^\#$ & \multicolumn{4}{c}{$\longleftarrow $
\qquad \quad $\ast {\color{dgray} ,Z_N^\#} \qquad \quad \longrightarrow$} & $\!\! \cdots$ \\[9pt]
$2$ & $Z_1^- \color{dgray} ,Z_1^\#$ & $Z_3^+$ & $\ \ast$ & $\ \ast$ & $\ \ast$ & $\ast$ \\[9pt]
$3$ & $Z_1 \color{dgray} ,Z_1^\#$  & \ccg $Z_3^-$ & \ccg $Z_5^+$ & \ccg $\ \ast$  & \ccg  $\ \ast$  & \ccg $\ast$ \\[9pt]
$4$ & $Z_1^+ \color{dgray} ,Z_1^\#$ &  \ccg $\ \ast$  & \ccg $Z_5^-$ & \ccg $Z_7^+$ &  \ccg  $\ \ast$  & \ccg $\ast$ \\[9pt]
$5$ & $Z_1^- \color{dgray} ,Z_1^\#$ & \ccg $Z_3$  &  \ccg  $\ \ast$  & \ccg $Z_7^-$ & \ccg  $Z_9^+$ & \ccg $\ast$ \\[4pt]
$6$ & $Z_1 \color{dgray} ,Z_1^\#$  & \ccg  $\ \ast$  &  \ccg  $\ \ast$  &  \ccg  $\ \ast$  & \ccg $Z_9^-$ & \ccg $\!\! \ddots$ \\[4pt]
$7$ & $Z_1^+ \color{dgray} ,Z_1^\#$ & \ccg $Z_3^+$ & \ccg $Z_5$  & \ccg  $\ \ast$  & \ccg  $\ \ast$  & \ccg $\!\! \ddots$ \\[6pt]
$\,\vdots$ & $\ \vdots$ &\ccg $\ \vdots$ \ \rotatebox[origin=c]{-20}{$\! \ddots$} 
& \ccg $\ \vdots $ \quad \rotatebox[origin=c]{-10}{$\! \ddots$} & \ccg & \ccg & \ccg  $\!\! \ddots$ \\[1pt]
\hline
\end{tabular}
\end{table}

\end{document}